\documentclass[aps,prb,amsmath,amssymb,reprint,floatfix]{revtex4-2}

\usepackage[utf8]{inputenc}
\usepackage{graphicx}
\usepackage{dcolumn}
\usepackage{bm}
\usepackage[hidelinks]{hyperref}
\usepackage{newtxtext,newtxmath}

\begin{document}

\title{Superconductivity in the kagome Hubbard model under the flat-band-preserving disorder} 

\author{Jicheol Kim}
\author{Dong-Hee Kim}
\email{dongheekim@gist.ac.kr}
\affiliation{Department of Physics and Photon Science, Gwangju Institute of Science and Technology, Gwangju 61005, Korea}

\begin{abstract}
We investigate the disordered flat-band superconductivity within the attractive Hubbard model on the kagome lattice by contrasting the flat-band-preserving disorder [Phys. Rev. B 98, 235109 (2018)] with the random hopping disorder that breaks the flat-band degeneracy. Through Bogoliubov--de~Gennes mean-field calculations, we find that the superfluid weight is much more robust under the flat-band-preserving disorder, while the system eventually undergoes a transition to an insulator as disorder becomes strong enough. The almost linear interaction-dependence of the superfluid weight in the weak coupling limit found with the flat-band-preserving disorder confirms the persistent flat-band signature, whereas the exponential behavior of a dispersive-band character arises with the random hopping counterpart. In addition, in the exact diagonalization of the one-particle density matrix, we identify an occupation spectrum structure attributed to the flat-band states, demonstrating the connection between the resilient flat band and the enhanced robustness of superconductivity.
\end{abstract}

\maketitle

\section{Introduction}

Superconductivity in flat-band systems has attracted increasing attention because of its properties that are fundamentally distinguished from conventional systems with dispersive bands and the relevance of the phenomena
in two-dimensional materials~\cite{Torma2022, Leykam2018, Yu2025}. For a conventional \textit{s}-wave superconductor with attractively interacting particles on a single dispersive band, the Bardeen-Cooper-Schrieffer (BCS) theory predicts the critical temperature $T_c \propto \exp(-1/|U|\rho(E_F))$, where $\rho(E_F)$ and $U$ are the density of states at the Fermi level and and the effective attractive interaction, respectively. In contrast, in a flat-band model, the mean-field calculations revealed that $T_c \propto |U|$ for the Cooper pair formation~\cite{Heikkila2011, Khodel1990, Kopnin2011}. Despite the diverging effective mass of a particle in a flat band, the superfluid weight $D_s$ can be finite even in an isolated flat-band limit due to the interplay between interactions and multiband effects producing the quantum geometric contributions~\cite{Peotta2015, Liang2017, Torma2018, Huhtinen2022, Rossi2021, Peotta:ENFI25, Tam2024, Chen2024, Chen2025}.

Disorder is an important ingredient in examining the robustness and vulnerability of a superconducting state, and its breakdown due to disorder has been an intriguing subject of study for a long time~\cite{Gantmakher2010, SITbook, Sacepe2020}. The Anderson theorem states that the \textit{s}-wave superconductivity is robust against a weak nonmagnetic disorder that preserves the time-reversal symmetry~\cite{Anderson1959, Abrikosov1959}. Ma and Lee~\cite{Ma1985} argued that it also survives with the localization of single-particle states that can occur at weak disorder in low dimensions~\cite{Anderson1958}. However, stronger disorder would lead to spatial inhomogeneity in the pairing, which would eventually destroy superconductivity. For instance, in the two-dimensional Hubbard model with attractive interactions, which we consider here for a flat-band system in our present study, the superconductor-insulator transition was extensively studied for square lattices under random onsite potentials, revealing the emergence of inhomogeneity in the local pairing and characterizing transport properties and spectral features \cite{Trivedi1996, Scalettar1999, Ghosal1998, Ghosal2001, Dubi2007, Bouadim2011, Sakaida2013, Poduval2022}.

The Hubbard model, realized with ultracold Fermi gases loaded on tunable optical lattices, may offer a highly controllable platform for studying flat-band effects in superconductivity and magnetism \cite{Tasaki1998, Tamura2002, Noda2009, Iglovikov2014, Julku2016, Huhtinen2021}, as exemplified by the recent realization of ferrimagnetism in the Lieb lattice \cite{Lebrat2025}. Beyond the clean system, past theoretical studies have explored the influence of disorder on the Hubbard models across various types of disorder and lattice geometries accommodating flat bands \cite{Oliveira-Lima2020, Roy2020, Li2022, Kiran2024, Lau2022, Liang2023, Chan2025, Bouzerar2025, Kolar2025}. In particular, there are a few numerical \cite{Lau2022, Liang2023, Chan2025, Bouzerar2025} and analytical \cite{Kolar2025} studies that have recently examined the fate of the flat-band superconductivity under disorder.

The mean-field study reported a universal suppression of superfluid weight across various settings of band flatness and topology in the extended Kane-Mele model under random onsite potentials \cite{Lau2022}. The density matrix renormalization group studies on the quasi-one-dimensional Creutz lattice found that there exists a finite critical disorder strength for random onsite potentials at any interaction strength \cite{Chan2025} as well as for random Zeeman fields \cite{Liang2023}. The mean-field study on the Lieb lattice under offdiagonal disorder demonstrated more robust superconductivity at the band than at the dispersive band \cite{Bouzerar2025}. Most recently, the competition between interaband and interband localization functions in the robust superfluid weight was discussed analytically \cite{Kolar2025}. 

Although these previous studies agreed on robust superconductivity in disordered flat-band systems and emphasized the role of the flat band, it is still nontrivial to understand the interplay between the flat band, disorder, and interaction. In noninteracting systems, a small amount of disorder lifts the flat-band degeneracy, significantly changing the localization property \cite{Goda2006, Nishino2007, Chalker2010, Leykam2013, Leykam2017, Shukla2018a, Shukla2018b}. In this context, it may be still worthwhile to consider an exotic type of disorder that does not perturb the noninteracting flat-band states, which would provide an explicit way to reveal the flat-band effects on the robustness of superconductivity. Specific questions include how such a preserved flat band would reveal itself in disordered superconductivity and whether it could help resist the anticipated breakdown as disorder strength further increases.

In this paper, we address these questions by contrasting an artificial disorder
that preserves the degenerate flat-band states with one that breaks the flat-band degeneracy
within the attractive Hubbard model on the kagome lattice.
In particular, we investigate a system under the disorder designed
by Bilitewski and Moessner \cite{Bilitewski2018} that preserves
the noninteracting flat-band states at any disorder strength on the kagome lattice.
We compare this ``flat-band-preserving'' (FBP) disorder
with the random hopping (HOP) disorder that takes only the offdiagonal part from the FBP type
to break the flat-band degeneracy. 
Our approach may be compared to Ref.~\cite{Bouzerar2025} that introduced the offdiagonal disorder
preserving the bipartite character of the Lieb lattice, which, however, cannot be applied 
to the kagome lattice whose hopping structure is not bipartite.
While both demonstrate the role of the flat band through a comparative analysis,
here we examine the picture of the superconductor-insulator transition and clarify
the persistent flat-band signature from the interaction-strength dependence of the superfluid weight
at zero temperature.
In addition, using the one-particle density matrix (OPDM), we propose another tool 
to discuss the robustness of the flat-band effects.

Using the Bogoliubov--de~Gennes approach, we investigate various superconducting observables at zero temperature, including pairing amplitude, superfluid weight, and density of states, for a fixed particle density corresponding to the half-filled flat band of a clean system. On the other hand, using exact diagonalization on small clusters, we compute the occupation spectrum of the OPDM. It turns out that while the system eventually undergoes a transition to an insulator when disorder is sufficiently strong, the superconductivity is much more robust with the FBP disorder. We find that the flat-band signature persists in the superfluid weight in the FBP case, whereas in the HOP counterpart, it already exhibits a dispersive-band-like character at much weaker disorder. Also, we verify the robustness and vulnerability of the flat-band effects for each type of the disorders using the characteristic discontinuity identified in the OPDM spectrum.

This paper is organized as follows. In Sec.~\ref{sec:method}, we provide the model Hamiltonian and describe the detail of the flat-band-preserving and random hopping disorders. Our numerical methods and procedures are also briefly explained in this section. In Sec.~\ref{sec:result}, we present our main results, including the profiles of local pairing amplitudes and particle densities, the superfluid weight, the density of states, and the occupation spectrum of the one-particle density matrix. Finally, the summary and conclusions are given in Sec.~\ref{sec:conclusion}.

\section{Model and Methods} \label{sec:method}

\subsection{Flat-band-preserving and random hopping disorders}
\label{subsec:method_disorder}

\begin{figure}[t]
    \includegraphics[width=1.0\linewidth]{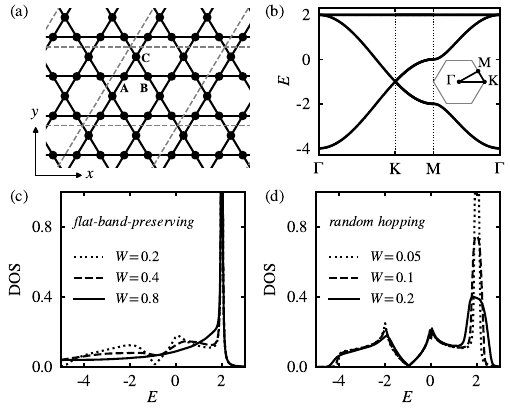}
    \caption{Flat band of the disordered kagome tight-binding model. (a) The structure of the kagome lattice. The dashed lines indicate the $L \times L$ supercells. The case of $L=2$ is exemplified. (b) The band structure for the clean system of the disorder strength $W = 0$. The density of states (DOS) is displayed for (c) the flat-band preserving disorder and (d) the random hopping disorder. The plots of DOS are made for $L = 128$ using the Lorentzian broadening of $0.01$ and averaged over $50$ different disorder realizations.}
    \label{fig1}
\end{figure}

The noninteracting tight-binding Hamiltonian with nonmagnetic disorder across bonds and sites can be written as
\begin{equation} \label{eq:H0}
\hat{H}_0 = - \sum_\sigma \sum_{\mathbf{r}_1,\mathbf{r}_2}
t_{\mathbf{r}_1, \mathbf{r}_2} \,
\left(
\hat{c}^{\dagger}_{\mathbf{r}_1 \sigma} \hat{c}_{\mathbf{r}_2 \sigma}
+ \mathrm{h.c.}
\right)
- \sum_{\mathbf{r}, \sigma} V_\mathbf{r} \, \hat{n}_{\mathbf{r} \sigma} \,,
\end{equation}
where the hopping strength $t_{\mathbf{r}_1, \mathbf{r}_2}$ is nonzero only for a bond between nearest-neighboring sites under periodic boundary conditions, and $V_\mathbf{r}$ is a onsite potential at site position $\mathbf{r}$. We consider a nonmagnetic system with $t_{\mathbf{r}_1, \mathbf{r}_2}$ and $V_\mathbf{r}$ being independent of spin component $\sigma \in \{ \uparrow, \downarrow \}$. For the kagome lattice in the clean limit ($t_{\mathbf{r}_1, \mathbf{r}_2} = t > 0$ and $V_\mathbf{r} = 0$), the flat band is located at the highest energy of the spectrum,
being touched by a dispersive band as displayed in Fig.~\ref{fig1}.

The flat-band-preserving (FBP) disorder for the kagome lattice designed by Bilitewski and Moessner \cite{Bilitewski2018} implements inhomogeneity into both of $t_{\mathbf{r}_1, \mathbf{r}_2}$ and $V_\mathbf{r}$ in the following way. First, a random number $\eta_{\mathbf{r}}$ is drawn for each site at $\mathbf{r}$ from a uniform distribution $\eta \in [1-W, 1+W]$. The disorder strength $W$ is tuned between $0$ and $1$. Then, these random numbers determine $t_{\mathbf{r}_1, \mathbf{r}_2}$ and $V_\mathbf{r}$ as
\begin{equation} \label{eq:FBP}
t_{\mathbf{r}_1, \mathbf{r}_2} = 
\begin{cases}
\eta_{\mathbf{r}_1} \eta_{\mathbf{r}_2} &
\text{for } (\mathbf{r}_1, \mathbf{r}_2) \in \mathcal{B} \\
0 & \text{otherwise}
\end{cases} 
\,,
\quad V_\mathbf{r} = 2 (\eta_\mathbf{r}^2 - 1)\,,
\end{equation}
where $\mathcal{B}$ indicates the site-pair set of bonds. In this particular design of disorder for the kagome lattice, the flat-band states in the tight-binding model are preserved at any disorder strength $W \in [0, 1)$, which we exemplify for selected values of $W$ in the density of states (DOS) shown in Fig.~\ref{fig1}(c).

On the other hand, as a counterpart to the FBP disorder, we need one that destroys the degeneracy of the flat-band states. We construct such disorder simply by switching off the onsite potentials ($V_\mathbf{r} = 0$) while keeping the same structure of the disordered hopping matrix. This ``random hopping'' (HOP) disorder lifts the flat-band degeneracy at $E = 2$ at any nonzero disorder strength $W$. As demonstrated in Fig.~\ref{fig1}(d), the peak in the density of states gets broadened as $W$ increases and is smeared away with stronger disorder. The advantage of choosing the HOP disorder, instead of more common random onsite potentials, is that by sharing the hopping matrix of the same random variable distribution between the FBP and HOP disorders, the effects of the two disorders can be compared in the same scale of disorder strength. The energy scale is naturally unified between the two by defining the energy unit as the average hopping strength, which is set to be unity as implied in Eq.~\eqref{eq:FBP} and omitted for brevity hereafter.

\subsection{Self-consistent mean-field approach}
\label{subsec:method_MFT}

We examine the disorder effects on the $s$-wave superconductivity within the attractive Hubbard model described as
\begin{equation} \label{eq:HubbardH}
\hat{H} = \hat{H}_0(\boldsymbol{\eta}) - \mu \sum_{\mathbf{r}, \sigma}
\hat{n}_{\mathbf{r} \sigma} - U \sum_\mathbf{r}
\hat{n}_{\mathbf{r}\uparrow} \hat{n}_{\mathbf{r}\downarrow} \,,
\end{equation}
where $U > 0$ is the onsite interaction strength and $\mu$ is the chemical potential. The noninteracting part $\hat{H}_0(\boldsymbol{\eta})$ for a disorder configuration $\boldsymbol{\eta}$ is given in Eq.~\eqref{eq:H0}. In this subsection, we briefly describe the self-consistent Bogoliubov--de~Gennes (BdG) mean-field method to solve this Hamiltonian.

We define the unit cell of the kagome lattice as a supercell of $L \times L$ primitive cells, which is illustrated in Fig.~\ref{fig1}(a). One unit cell accommodates a disorder configuration $\boldsymbol{\eta} \equiv (\eta_1, \eta_2, \ldots, \eta_N)$ of $N = 3L^2$ sites, periodically tiling the entire system. A site in the system is labeled with an index $l$ of the cell that accommodates the site and a sublattice index $\alpha$ pointing the site within the cell. The position vector of site $(l,\alpha)$ can be written as
$\mathbf{r}_{l,\alpha} = \mathbf{R}_l + \mathbf{r}_\alpha$, 
where $\mathbf{R}_l$ points to the origin of the cell from which the vector $\mathbf{r}_\alpha$ locates the site within the cell. We write the Fourier transformation as 
$\hat{c}_{\mathbf{k}\alpha\sigma} = N_c^{-1/2}
\sum_l \mathrm{e}^{-i\mathbf{k}\cdot \mathbf{r}_{l,\alpha}}\,
\hat{c}_{l\alpha\sigma}$,
where $N_c$ is the number of unit cells in the system, and the momentum $\mathbf{k}$ is restricted to the Brillouin zone. Using the mean-field approximation, one can write down the effective BdG Hamiltonian as
$\hat{H}_\mathrm{eff} = \sum_\mathbf{k} \hat{\Psi}^\dagger_\mathbf{k}
\mathbf{H}_\mathrm{BdG}(\mathbf{k}) \hat{\Psi}_\mathbf{k}$
with the Nambu spinor
$\hat{\Psi}_\mathbf{k} = (
\hat{c}_{\mathbf{k}1\uparrow}, 
\hat{c}_{\mathbf{k}2\uparrow}, 
\ldots,
\hat{c}_{\mathbf{k}N\uparrow}, 
\hat{c}^\dagger_{\mathbf{-k}1\downarrow}, 
\hat{c}^\dagger_{\mathbf{-k}2\downarrow}, 
\ldots,
\hat{c}^\dagger_{\mathbf{-k}N\uparrow}
)^T$.
The matrix $\mathbf{H}_\mathrm{BdG}(\mathbf{k})$ 
can be diagonalized as
\begin{equation} \label{eq:bdg_matrix}
\begin{pmatrix}
   \mathbf{h}_\mathbf{k}
   - \boldsymbol{\lambda} & \boldsymbol{\Delta} \\ 
   \boldsymbol{\Delta}^\dagger &
   -\mathbf{h}_\mathbf{k} + \boldsymbol{\lambda}
\end{pmatrix}
\begin{pmatrix}
    \mathbf{u}_{n\mathbf{k}} \\
    \mathbf{v}_{n\mathbf{k}}
\end{pmatrix}
= 
E_{n\mathbf{k}}
\begin{pmatrix}
    \mathbf{u}_{n\mathbf{k}} \\
    \mathbf{v}_{n\mathbf{k}} 
\end{pmatrix}
\,,
\end{equation}
where $\mathbf{h}_\mathbf{k}$ is the kinetic part of the Hamiltonian, given as
\begin{equation}
[\mathbf{h}_\mathbf{k}]_{\alpha \beta} = -\sum_l t_{\mathbf{r}_{0, \alpha}, \mathbf{r}_{l, \beta}}\, \exp[-i \mathbf{k} \cdot (\mathbf{r}_{0,\alpha} - \mathbf{r}_{l, \beta})] \,.
\end{equation}
The relation $\mathbf{h}_{\uparrow,\downarrow}(\mathbf{k}) = \mathbf{h}^\ast_{\downarrow,\uparrow}(-\mathbf{k}) \equiv \mathbf{h}_\mathbf{k}$ has been used for the time-reversal symmetry in our spin-unpolarized system. The element of the diagonal matrix $\boldsymbol{\Delta}$ is the local pairing amplitude $\Delta_\alpha = -U / N_c \sum_\mathbf{k} \langle \hat{c}_{-\mathbf{k}\alpha\downarrow} \hat{c}_{\mathbf{k}\alpha\uparrow} \rangle$, 
and the diagonal matrix $\boldsymbol{\lambda}$ has an element $\lambda_\alpha = V_\alpha + U n_\alpha /2 + \mu$ with $n_\alpha = \langle \hat{n}_{\alpha\uparrow} \rangle + \langle \hat{n}_{\alpha\downarrow} \rangle$. The term $U n_\alpha / 2$ indicates the site-dependent Hartree shift to be determined self-consistently. The inclusion of the Hartree shift increases a numerical complexity but is crucial for describing inhomogeneity emerging in a disordered superconductor.

The local pairing amplitude $\{\Delta_\alpha\}$ and the site occupancy $\{n_\alpha\}$ can be computed at zero temperature as 
\begin{eqnarray}
    \Delta_\alpha &=& \frac{U}{N_c}
    \sum_\mathbf{k} \sideset{}{'}\sum_n
    [\mathbf{u}_{n\mathbf{k}}]_\alpha
    [\mathbf{v}^\ast_{n\mathbf{k}}]_\alpha \,, \\
    n_\alpha &=& \frac{2}{N_c}
    \sum_\mathbf{k} \sideset{}{'}\sum_n
    \big\vert [\mathbf{v}_{n\mathbf{k}}]_\alpha \big\vert^2
    \,,
\end{eqnarray}
where the spin-density balance is imposed, and the primed sum indicates partial summation over the eigenstates with $E_n > 0$. Subject to the fixed density $\bar{n} \equiv \sum_\alpha n_\alpha / N = 5/3$, which corresponds to the half-filled flat band in the clean limit, we iteratively solve the BdG equations to determine  $\{\Delta_\alpha\}$, $\{n_\alpha\}$, and $\mu$ self-consistently. Additionally, we compute the local density of states $\mathcal{N}_\alpha(\omega)$, which is written as 
\begin{eqnarray}
    \mathcal{N}_\alpha(\omega) = \frac{1}{N_c} \sum_\mathbf{k}
    \sideset{}{'}\sum_{n}
    &\Big[& \big|[\mathbf{u}_{n\mathbf{k}}]_\alpha \big|^2
    \delta(\omega - E_{n\mathbf{k}}) \\ \nonumber
    &+& \big| [\mathbf{v}_{n\mathbf{k}}]_\alpha \big|^2
    \delta(\omega + E_{n\mathbf{k}})\,\Big] \,,
\end{eqnarray}
where $\delta(x)$ denotes the Dirac delta function. The density of states is given as  $\mathcal{N}(\omega) = \sum_\alpha \mathcal{N}_\alpha(\omega) / N$. For visualization, we approximate the delta function by the Lorentzian shape with the broadening factor of $0.01$ in the energy unit.

In practice, we solve the BdG equations mostly in the real-space representation, where we assume that the system is a cluster of one large unit cell ($N_c = 1$ and $\mathbf{k} = 0)$. In this real-space BdG calculation, one can access a large $L$ within limited computational time, which is often preferred to reduce the finite-size effects in disordered systems. However, to extract the geometrical contribution from the superfluid weight, the momentum-space formulation \cite{Peotta2015, Liang2017, Lau2022} is essential. Although a practically achievable $L$ was turned out to be smaller in our momentum-space calculations, we have checked consistency between the real-space and momentum-space calculations in the estimate of the total superfluid weight.

\subsection{Superfluid weight}
\label{subsec:method_superfluid_weight}

In this subsection, we review the real-space \cite{Ghosal1998, Ghosal2001, Kiran2024} and momentum-space \cite{Peotta2015, Liang2017} BdG calculations of the superfluid weight. For the real-space calculation in the kagome lattice, we follow the formulation given in Ref.~\cite{Kiran2024}. In the linear response theory with the Kubo formula \cite{Scalapino1993}, the superfluid weight is written as
\begin{equation} \label{eq:Ds1}
    D_s = \langle -\hat{k}_x \rangle
    - \Lambda_{xx}(q_x = 0, q_y \to 0, i\omega_n = 0). 
\end{equation}
While this expression is for the $xx$-component, we will simply call it as the superfluid weight $D_s$ because of the rational symmetry \cite{Liang2017, Kiran2024}. Disorder breaks the rotational symmetry of the kagome lattice, but the symmetry is recovered after disorder averaging. The first term is the diamagnetic response given by  the kinetic energy density in the $x$ direction, 
\begin{equation}
\langle -\hat{k}_x \rangle = \frac{4}{A}
\sum_{\alpha} \sum_{\beta \in \mathcal{X}^+_\alpha}
\sideset{}{'}\sum_n 
D_{\alpha\beta}\, t_{\alpha \beta} \, v_{n\alpha} v_{n\beta}\,,
\end{equation}
where $\mathcal{X}^+_\alpha$ is a set of the nearest neighbors of site $\alpha$ that reside in the right-hand side of site $\alpha$ subject to the periodic boundary conditions, and $D_{\alpha\beta} = 1$ and $1/4$ for horizontal (A--B) and oblique (A--C or B--C) bonds (see Fig.~\ref{fig1}), respectively. The area $A$ is set to be $2 \sqrt{3} L^2$ for the cluster of $L\times L$ primitive cells by taking the bond length equal to one. The notations of the BdG eigenvector components are $u_{n\alpha} \equiv [\mathbf{u}_{n,\mathbf{k}=0}]_\alpha$ and $v_{n\alpha} \equiv [\mathbf{v}_{n,\mathbf{k}=0}]_\alpha$. The second term representing the paramagnetic response can be evaluated as
\begin{eqnarray} \label{eq:Lxx}
    \Lambda_{xx}(\mathbf{q}, 0) &=& \frac{2}{A}
    \sum_{\alpha, \alpha^\prime}
    \mathrm{e}^{i\mathbf{q} \cdot (\mathbf{r}_\alpha - \mathbf{r}_{\alpha^\prime})}
    \sum_{\beta \in \mathcal{X}^+_\alpha} 
    P_{\alpha \beta}
    \sum_{\beta^\prime \in \mathcal{X}^+_{\alpha^\prime}} 
    P_{\alpha^\prime \beta^\prime}
    \nonumber \\
    && \times \sideset{}{'}\sum_{n,m} 
    \frac{t_{\alpha\beta} t_{\alpha^\prime \beta^\prime}}{E_n + E_m} 
    \Big[
        \left(
            u_{n\alpha^\prime} v_{m\beta^\prime}
            + v_{n\beta^\prime} u_{m\alpha^\prime} 
        \right) 
    \nonumber \\
    && \times
        \left(
            u_{n\beta} v_{m\alpha}
            - v_{n\beta} u_{m\alpha}
            - u_{n\alpha} v_{m\beta}
            + v_{n\alpha} u_{m\beta}
        \right)
    \nonumber \\
    && + \left( u \leftrightarrow v \right) \Big]\,,
\end{eqnarray}
where $P_{\alpha\beta}$ is $1$ for the horizontal bonds and $1/2$ otherwise. 

On the other hand, the superfluid weight is separated into the conventional and geometric contributions in the momentum-space expression \cite{Peotta2015,Liang2017}. Using the symmetry relation of $\mathbf{h}_\mathbf{k}$, the superfluid weight can be written as
\begin{equation} \label{eq:Ds2} 
[D_{s}]_{\mu\nu} = \frac{1}{A} \sum_{\mathbf{k}}
\sum_{r, s, p, q} 
C^{rs}_{pq}(\mathbf{k}) 
[j_\mu(\mathbf{k})]_{rs}
[j_\nu(\mathbf{k})]_{pq}\,,
\end{equation}
where the diagonal partial sum with $r = s$ and $p = q$ identifies the conventional contribution $D_s^\mathrm{conv}$, and then the geometric contribution can be obtained as $D_s^{\mathrm{geom}} = D_s - D_s^{\mathrm{conv}}$. The matrix elements of the current operator are given as
\begin{eqnarray} \label{eq:jx}
    [j_\mu(\mathbf{k})]_{rs} = 
    \mathbf{w}^\dagger_{r\mathbf{k}}
    (\partial_{k_\mu}\mathbf{h}_\mathbf{k})\,
    \mathbf{w}_{s\mathbf{k}} \,,
\end{eqnarray}
where $\mathbf{w}_{q\mathbf{k}}$ is the $q^\mathrm{th}$ eigenvector of the matrix $\mathbf{h}_\mathbf{k} - \boldsymbol{\lambda}$, generating an orthonormal basis set. The coefficient $C^{rs}_{pq}(\mathbf{k})$ is evaluated for a given $\mathbf{k}$ as
\begin{equation} \label{eq:Ctensor}
    C^{rs}_{pq} =
    4 \sum_{m,n} \frac{n_\mathrm{F}(E_m) - n_\mathrm{F}(E_n)}{E_m - E_n} 
    \tilde{u}^\ast_{mr} \tilde{u}_{ns} \tilde{v}^\ast_{np} \tilde{v}_{mq}\,,
\end{equation}
where the momentum $\mathbf{k}$ is dropped for the brevity of expressions, and the Fermi-Dirac distribution $n_F(E)$ is evaluated in the zero-temperature limit. The elements of $\{\tilde{u}\}$ and $\{\tilde{v}\}$ are the coefficients appearing in the linear transformation of the BdG eigenvector $(\mathbf{u}_{n\mathbf{k}}, \mathbf{v}_{n\mathbf{k}})^T$ from the site basis $\{\alpha\}$ to the eigenbasis $\{\mathbf{w}_{q\mathbf{k}}\}$ of the matrix $\mathbf{h}_\mathbf{k} - \boldsymbol{\lambda}$ and thus can be obtained by solving the linear equations,
\begin{equation} \label{eq:coeff}
\mathbf{u}_{n\mathbf{k}} = \sum_{q} \tilde{u}_{nq} \mathbf{w}_{q\mathbf{k}}, \quad
\mathbf{v}_{n\mathbf{k}} = \sum_{q} \tilde{v}_{nq} \mathbf{w}_{q\mathbf{k}}.
\end{equation}
Note that in Eqs.~\eqref{eq:Ds2}--\eqref{eq:coeff}, have used the symmetry relation $\mathbf{h}_\mathbf{k} \equiv \mathbf{h}_{\uparrow,\downarrow}(\mathbf{k}) = \mathbf{h}^\ast_{\downarrow,\uparrow}(-\mathbf{k})$  and the corresponding symmetry in the BdG eigenpairs: if $(\mathbf{u}, \mathbf{v})^T$ is the eigenvector with eigenvalue $E$, then $(\mathbf{v}, -\mathbf{u})^T$ is the eigenvector with eigenvalue $-E$, and vice versa, which leads to $C^{rs}_{pq} = C^{pq}_{rs}$.

Our BdG estimate of the superfluid weight does not consider the response of the pairing amplitude  $\boldsymbol{\Delta}$ to the vector potential, which has recently been proposed for multiband \cite{Huhtinen2022, Tam2024} and disordered \cite{Schirmer2025, Kolar2025} systems. In our sample test of the method proposed in Ref.~\cite{Schirmer2025}, the overestimate was about $5$--$10\%$, but it did not change the qualitative behavior that we discuss in the next section. The other important source of overestimation is the quantum fluctuations neglected in the mean-field framework \cite{Ghosal1998, Ghosal2001}, which has not been tested so far in disordered flat-band systems. These prevent us locating a precise critical point of the superconductor-insulator transition. In the present work, we instead focus on a qualitative analysis of the flat-band effects on disordered superconductivity by comparing the two different types of disorder. 

\subsection{One-particle density matrix}

\begin{figure}
    \includegraphics[width=1.0\linewidth]{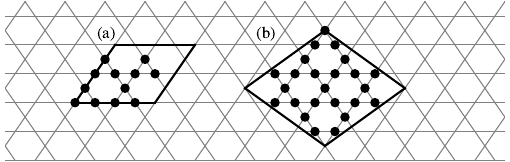}
    \caption{Geometry of the clusters with (a) $N=12$ and (b) $N=24$ sites used for the exact diagonalization calculation of the one-particle density matrix and the occupation spectrum.}
    \label{fig2}
\end{figure}

The one-particle density matrix $\boldsymbol{\rho}^{(1)}$ (OPDM) \cite{Penrose1956} is written for the ground state in our system as
\begin{equation} \label{eq:opdm}
    \rho^{(1)}_{ij} =  \langle \Psi_\mathrm{g} |
    \hat{c}^\dagger_{i\uparrow} \hat{c}_{j\uparrow}
    | \Psi_\mathrm{g} \rangle = \langle \Psi_\mathrm{g} |
    \hat{c}^\dagger_{i\downarrow} \hat{c}_{j\downarrow}
    | \Psi_\mathrm{g} \rangle \,,
\end{equation}
where $i$ and $j$ are the site indices, and $|\Psi_\mathrm{g}\rangle$ is the ground-state many-body wavefunction. The usefulness of the information extracted from the OPDM has been demonstrated in various quantum many-body systems under disorder or inhomogeneity, including hard core bosons in a trap and a quasi-periodic potential \cite{Rigol2004, Rigol2005, Nessi2011, Gramsch2012} and many-body localized systems \cite{Bera2015, Bera2017, Orito2021, Hopjan2021}. 

The eigenproblem, $\boldsymbol{\rho}^{(1)} |\tilde{\alpha}\rangle = \tilde{n}_{\tilde{\alpha}} |\tilde{\alpha}\rangle$, provides an occupation number $\tilde{n}_{\tilde{\alpha}}$ for an eigenstate $\tilde{\alpha}$ referred to as a natural orbital. In the occupation spectrum, we identify the structure linked to the flat-band states and see how it evolves with increasing disorder strength. While we examine the BdG mean-field wave function as well for the OPDM, our analysis of the OPDM spectrum on the robustness of the flat-band features is mainly based on the exact ground-state wave function for the small clusters shown in Fig.~\ref{fig2}.

\section{Results and Discussions}
\label{sec:result}

All results presented in this section are computed with a fixed particle density $\bar{n} \equiv \sum_\sigma \langle \hat{n}_\sigma \rangle / N$ at $\bar{n} = 5/3$, which corresponds to the half-filled flat band in the clean kagome lattice. The results are averaged over at least $100$ random disorder configurations. The error bars, which we define as the sample-to-sample fluctuations in this work, are omitted unless they are larger than the marker size.

\subsection{Superconductor-insulator transition}
\label{sec:result_sit}

Let us first present the disorder-strength dependence of  the superconducting observables obtained from the real-space BdG calculations at the flat-band half-filling in the kagome-lattice. In the dispersive-band case of the square lattice under onsite random potentials, it is well known that the system undergoes a transition from the superconducting phase to the insulating phase as the disorder strength increases \cite{Ghosal1998, Ghosal2001}. The main question that we want to address in this subsection is whether or not the flat-band-preserving (FBP) disorder changes the picture of the superconductor-insulator transition. In the noninteracting system, the FBP disorder preserves the flat-band degeneracy at any strength of disorder, from which it could be anticipated that the flat-band superconductivity might be very robust under this particular type of disorder. The subsequent question is then how much the robustness differs between the FBP disorder and the random hopping (HOP) counterpart that destroys the flat-band degeneracy.

\begin{figure}
    \includegraphics[width=1.0\linewidth]{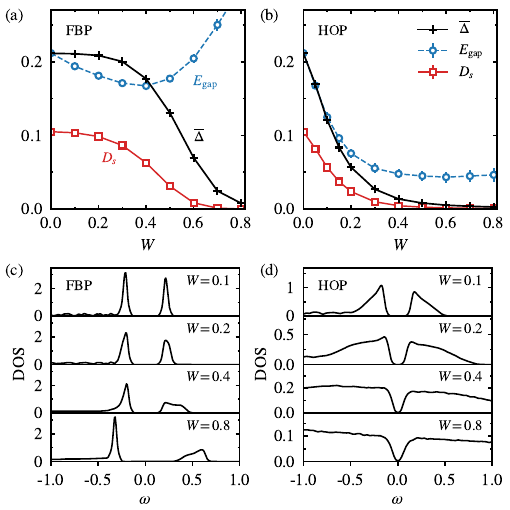}
    \caption{Superconductor-insulator transition. The site-averaged pairing amplitude $\overline{\Delta}$, the superfluid weight $D_s$, and the energy gap $E_\mathrm{gap}$ are plotted as a function of disorder strength $W$ and compared between (a) the flat-band-preserving (FBP) disorder and (b) the random hopping (HOP) disorder. The panels (c) and (d) display the single-particle density of states (DOS) for selected values of $W$. The real-space BdG calculations are performed in the system of size $L = 24$ ($1728$ sites) with the interaction strength $U = 1$.}
    \label{fig3}
\end{figure}

\begin{figure}
    \includegraphics[width=1.0\linewidth]{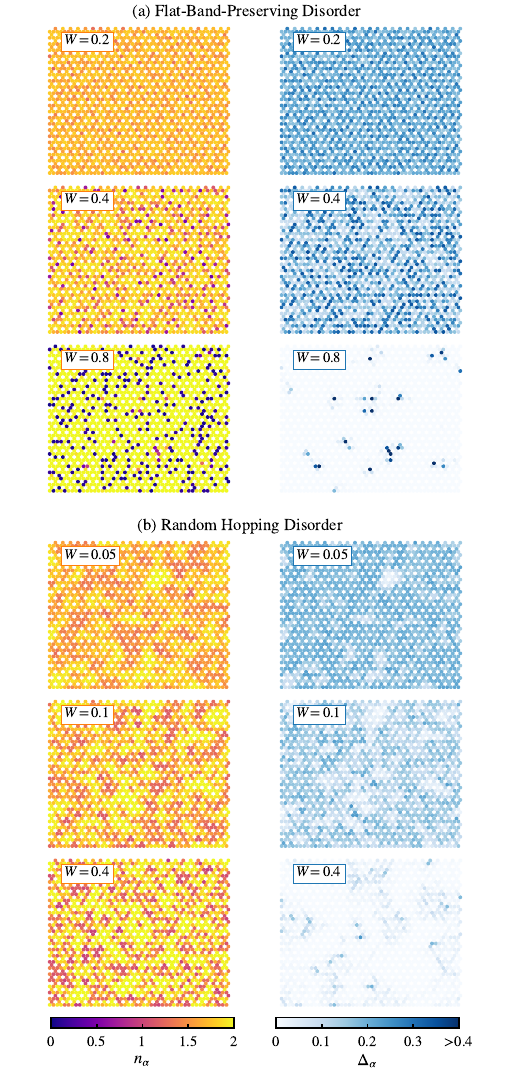}
    \caption{Site occupancy (left) and local pairing amplitude (right). The sample profiles are collected for selected values of disorder strength $W$ using the BdG calculations with $U = 1$ in the system of $L=24$.}
    \label{fig4}
\end{figure}

Figure~\ref{fig3} displays our real-space BdG calculation results for $U = 1$. It turns out that both types of disorder exhibit the characteristics of the superconductor-insulator transition, while the critical disorder is not located precisely within our mean-field calculations. The superfluid weight becomes suppressed with strong enough disorder, leading to the breakdown of superconductivity, even for the flat-band-preserving case. The FBP disorder is not immune to the spatial fluctuations as visualized in Fig.~\ref{fig4}. The density fluctuations lead to the spatial inhomogeneity in the Hartree field, which works as an additional disorder field destroying the degeneracy of the noninteracting flat-band states. Our numerical calculations suggest that such spatial perturbation becomes relevant to the decay of superconductivity for the disorder strength $W \gtrsim 0.2$ in the case of the FBP disorder.

On the other hand, we observe that the energy gap in the quasiparticle excitation spectrum never decreases below a certain value, even at a high disorder strength where the superfluid weight is almost suppressed. This agrees with the known feature of the insulating phase in the superconductor-insulator transition \cite{Ghosal1998, Ghosal2001}.
In the clean limit, the energy gap $E_\mathrm{gap}$ is equivalent to the average pairing amplitude $\bar\Delta = \sum_\alpha \Delta_\alpha / N$, and both quantities decrease with increasing $W$. However, at a certain disorder strength, $E_\mathrm{gap}$ turns to increase, while $\bar\Delta$ keeps decreasing. Also, in the case of the HOP disorder, $E_\mathrm{gap}$ remains significant at strong disorder, although the increasing behavior is not as pronounced as in the FBP case. The similar features can also be found in the single-particle density of states, where the gap is nonvanishing while the coherence peaks are smeared out at strong disorder.

Another important feature expected for strong disorder is the emergence of superconducting islands, which leads to a finite $\bar\Delta$ in the insulating phase. Figure~\ref{fig4} presents the profiles of the site occupancy $n_\alpha$ and the local pairing amplitude $\Delta_\alpha$ for various disorder strengths, displaying the formation of superconducting islands with $\Delta_\alpha \neq 0$ surrounded by insulating sites with $\Delta_\alpha = 0$. In a closer look, those insulating sites are mostly either fully occupied or empty in our kagome system, and the less common partially occupied sites in this area exhibit a gap in the local density of states. 

The picture of the superconductor-insulator transition does not depend on the type of disorder that we have examined. However, a contrast between the two is pronounced in the weakly disordered regime. In the case of the FBP disorder, the superfluid weight and the average pairing amplitude are almost unchanged for $W \lesssim 0.2$. In this range of $W$, the density fluctuations seems to be irrelevant, which would lead to robust flat-band effects as intended with the FBP disorder. In contrast, under the HOP disorder, which does not preserve the flat band, a significant decrease in the superconducting observables starts at a much lower disorder strength.

\subsection{Flat-band signature in superfluid weight}
\label{sec:result_ds}

\begin{figure}
    \centering
    \includegraphics[width=1.0\linewidth]{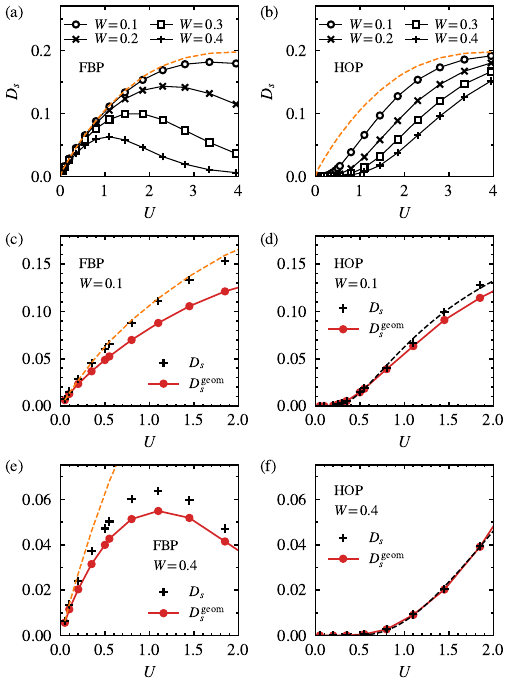}
    \caption{Superfluid weight as a function of interaction strength $U$. Left and right panels correspond to the flat-band-preserving (FBP) an random hopping (HOP) disorders, respectively. In (c)--(f), the geometric contribution $D_s^\mathrm{geom}$ is compared with the total superfluid weight $D_s$ for the selected values of disorder strength, $W=0.1$ and $W = 0.4$. The dashed lines in (a)--(c) and (e) indicate the superfluid weight for the clean system ($W = 0$). The dashed lines in (d) and (f) are the curve fits to the form of $a\exp(-b/U)$. The momentum-space BdG calculations are performed in the system of $L = 6$ ($108$ sites).}
    \label{fig5}
\end{figure}

The above comparisons suggest a connection between the preserved flat band and the enhanced robustness of superconducting observables. However, more direct evidence is required to confirm that the disordered superconductivity is indeed of the flat-band type. We demonstrate this through the interaction-strength dependence of the superfluid weight. A signature of the flat-band superconductivity is the linear dependence, $D_s \propto U$, in the weak coupling limit, which was rigorously proven in the isolated flat-band limit~\cite{Peotta2015}. 
For a flat band touching a dispersive band, as seen in the kagome lattice, previous works reported the similar behavior with a logarithmic correction, $D_s \propto U\ln(a/U)$, in two dimensions \cite{Julku2016, Iskin2019, Wu2021, Huhtinen2022} and the sublinear behavior, $D_s \propto U^\varphi$, in quasi-one dimension \cite{Chan2022}.
Figure~\ref{fig5} shows that the characteristic linear behavior with a correction is noticeable only in the FBP case; in the case of the HOP disorder, the exponential behavior is observed for the examined values of disorder strengths, indicating the character of a dispersive band.

In the FBP case, we find that the overall shape of $D_s(U)$ in the clean limit is essentially retained under the disorder: it increases almost linearly for small $U$, reaches the maximum, and then decreases as $U$ further increases. This behavior was also discussed in the gapless Lieb lattice \cite{Julku2016}. As the disorder strength increases, the maximum and its location are progressively lowered. In addition, a large geometric contribution $D_s^\mathrm{geom}$ is observed as expected for the flat band, while the conventional contribution $D_s^\mathrm{conv}$ remains finite under the disorder. From the comparison with the HOP case that we continue to discuss below, this is possibly related to the system-spanning noncontractible loop states, created by the band touching, that are preserved under the FBP disorder \cite{Bilitewski2018}.

On the other hand, for the HOP disorder that perturbs the noninteracting flat-band states, we observe that $D_s(U)$ agrees better with the exponential form of the usual BCS order parameter, $D_s(U) \propto \exp(-b/U)$, expected in a dispersive-band system. This implies that superconductivity under the HOP disorder, while it still survives at weak disorder, may have already lost its flat-band character. It is worth noting that under the HOP disorder, the geometric contribution becomes increasingly dominant as disorder strength $W$ increases: for instance, $D_s \simeq D_s^\mathrm{geom}$ at $W = 0.4$ as shown in Fig.~\ref{fig5}(f). This is in contrast to the FBP case exhibiting a finite $D_s^\mathrm{conv}$, suggesting a necessity of further study on the relation to the destruction
of the system-spanning modes.

\subsection{One-particle density matrix}
\label{sec:result_opdm}

\begin{figure}
    \includegraphics[width=1.0\linewidth]{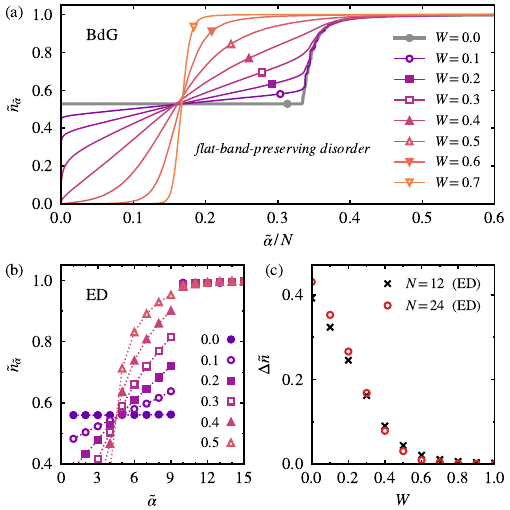}
    \caption{Occupation spectrum for the flat-band-preserving disorder. The one-particle density matrix is computed at $U=1$ for various values of disorder strength $W$. (a) BdG calculations of the occupation number $\tilde{n}_{\tilde{\alpha}}$. The system sizes examined are $L = 24$ ($1728$ sites) for $W \neq 0$ and $L = 48$ for $W = 0$. (b) Exact diagonalization (ED) results for the $24$-site cluster. (c) The jump $\Delta\tilde{n} \equiv \tilde{n}_{N/3+2} - \tilde{n}_{N/3+1}$ is plotted as a function of disorder strength with the ED calculations for the two different sizes of the cluster.}
    \label{fig6}
\end{figure}

Finally, we present the calculation of the occupation spectrum of the one-particle density matrix (OPDM), defined in Eq.~\eqref{eq:opdm}. We identify a flat-band feature in the occupation spectrum and analyze how it changes with increasing disorder strength. We then compare the robustness of the flat-band feature between the cases with the FBP and HOP disorders. The OPDM has been previously used to characterize the many-body localization in disordered interacting systems \cite{Bera2015, Bera2017, Orito2021, Hopjan2021}. However, to our knowledge, it has not been applied to a disordered superconductor with a flat band.

Let us first present the occupation spectrum in the clean system. The spectrum consists of the eigenvalue $\tilde{n}_{\tilde{\alpha}}$ of the OPDM associated with the natural orbital $\tilde{\alpha}$, sorted in the ascending order as $\tilde{n}_1 \le \tilde{n}_2 \le \ldots \le \tilde{n}_N$. As demonstrated in Fig.~\ref{fig6}, both of the real-space BdG method and the exact diagonalization (ED) reveal a plateau of a constant $\tilde{n}_{\tilde{\alpha}}$. It turns out that $N/3 + 1$ natural orbitals participate in this plateau in the system of $N$ sites. This is exactly the same number as in the flat band, which is spanned by $N/3-1$ compact localized states and two noncontractible loop states due to the singular band touching \cite{Bergman2008, Rhim2019, Rhim2021}. The precise agreement in the number of states raises a possibility that the noninteracting description of the flat band using the compact localized states may still provide an effective picture within the OPDM in the interacting model for the flat-band superconductivity.

\begin{figure}
    \includegraphics[width=1.0\linewidth]{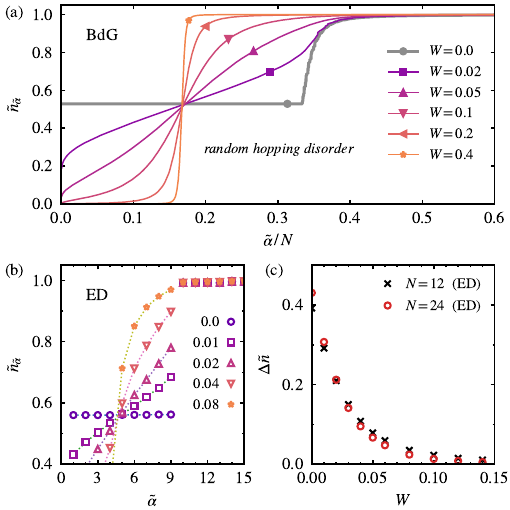}
    \caption{Occupation spectrum for the random hopping disorder. The system settings and the plot descriptions are identical to Fig.~\ref{fig6}.} 
    \label{fig7}
\end{figure}

The plateau ends with a sudden change at the edge, indicating a spectral discontinuity between the plateau and the other natural orbitals (see Figs.~\ref{fig6} and \ref{fig7} for $W=0$), although there is a difference between the BdG mean-field and ED calculations. In the BdG calculation, the kink at the edge looks like a discontinuity in the slope. However, the mean-field theory lacks quantum fluctuations, which may have affected the appearance of the kink particularly from the side that corresponds to a dispersive band. On the other hand, the ED shows a sizable discontinuity in the occupation number, which can be measured by the jump $\Delta \tilde{n} \equiv \tilde{n}_{N/3+2} - \tilde{n}_{N/3+1}$ between two neighboring natural orbitals at the edge. While the small accessible system sizes prevent us from doing a proper finite-size scaling analysis, we do not see any significant change in $\Delta \tilde{n}$ between the systems of $N = 12$ and $N = 24$ sites. While detailed characterization of the discontinuity may require more advanced theoretical methods, below we decide to examine the behavior of the plateau
and the jump using the ED to discuss the influence of disorder.

We find that disorder breaks the degeneracy of the plateau, irrespective of the type of disorder, leading to a finite slope in the spectrum. As displayed in Figs.~\ref{fig6} and \ref{fig7} for various values of disorder strength, the spectrum changes from the plateau in the clean limit to a structure resembling the Fermi-Dirac distribution in the strong disorder limit. Remarkably, below a certain strength of disorder, a finite jump exists at the edge, separating the first $N/3 + 1$ natural orbitals in the previously plateau region from the remainder of the spectrum. Assuming a connection between the flat-band states and the separated natural orbitals of the same quantity,
we may be allowed to use the nonvanishing discontinuity as an empirical indicator of persistent flat-band effects under disorder.

Measuring $\Delta \tilde{n}$ as a function of disorder strength $W$, we revisit the occupation spectrum for the FBP and HOP disorders. It turns out that the disorder-strength scale of the decay in $\Delta\tilde{n}$ is drastically different between the two types of disorder. In the case of the FBP disorder, $\Delta\tilde{n}$ remains finite until $W \sim 0.5$, whereas in the HOP case, it vanishes at a much lower strength of $W \sim 0.1$. This comparison of the suppression of $\Delta\tilde{n}$ with increasing $W$ in the OPDM spectrum shows qualitative agreement with the behavior of the superfluid weights contrasted between the FBP and HOP disorders, strengthening the evidence for the role of the resilient flat band in the enhanced robustness of superconductivity against disorder.

\section{Summary and Conclusions}
\label{sec:conclusion}

We have demonstrated that a preserved flat band induces more robust superconductivity under disorder within the attractive Hubbard model on the kagome lattice by contrasting the flat-band-preserving (FBP) disorder \cite{Bilitewski2018} and the random hopping (HOP) counterpart that breaks the flat-band degeneracy. Through Bogoliubov-de Gennes mean-field calculations, we have observed significantly more robust superconducting properties under FBP disorder, although the system eventually undergoes a superconductor-insulator transition at strong enough disorder. Analyzing the interaction-strength dependence of the superfluid weight, we have found that the FBP disorder indeed exhibits the signature of the flat-band superconductivity, showing a contrast with the HOP case where the exponential behavior expected in a dispersive-band system appear in the superfluid weight at much weaker disorder.

Additionally, we have proposed a spectral structure linked to the flat-band states in the occupation spectrum of the one-particle density matrix (OPDM). In the clean limit, the plateau of natural orbitals isolated by an edge discontinuity corresponds to exactly the same number of the flat-band states. Using the exact diagonalization method, we have attempted to connect the behavior of the occupation jump at the edge of the plateau region to the robustness of flat-band effects with increasing disorder strength. We have observed a significant decay of the jump with weak HOP disorder, whereas in the case of the FBP disorder, a finite jump is preserved even at a much higher disorder strength. These observations suggest that the OPDM can be another useful tool for studying interacting flat-band systems under perturbation.

While our observation suggests a connection between the flat-band states and the discontinuous plateau identified in the OPDM spectrum, further study is required to rigorously prove or examine the proposed connection across a broader range of interacting flat-band systems. For future investigations on the kagome lattice, specific questions can include whether the compact localized states and the noncontractible loop states can still form a basis set for the natural orbitals associated with the discontinuous plateau region. A related but more general question can be whether and how a perturbation deforms the compact localized states in interacting systems.

\begin{acknowledgments}
We thank Alexei Andreanov and Sebastiano Peotta for insightful discussions. 
This work was supported by the Gwangju Institute of Science and Technology (GIST) research fund (Future-leading Specialized Research Project, 2025) and the Regional Innovation System \& Education(RISE) program through the (Gwangju RISE Center), funded by the Ministry of Education(MOE) and the (Gwangju Metropolitan City), Republic of Korea (2025-RISE-05-001).
\end{acknowledgments}

\bibliography{paper}

@article{Torma2022,
  title = {Superconductivity, superfluidity and quantum geometry in twisted multilayer systems},
  author = {T\"orm\"a, P\"aivi and Peotta, Sebastiano and Bernevig, Bogdan A.},
  journal = {Nat. Rev. Phys.},
  volume = {4},
  pages = {528--542},
  year = {2022},
  doi = {10.1038/s42254-022-00466-y}
}

@article{Heikkila2011,
  title ={Flat bands in topological media},
  author = {Heikkil\"{a}, T. T. and Kopnin, N. B. and Volovik, G. E.},
  journal = {JETP Lett.},
  volume = {94},
  pages = {233},
  year = {2011},
  doi = {10.1134/S0021364011150045}
}

@article{Khodel1990,
  title = {Superfluidity in systems with fermion condensate},
  author = {Khodel', V. A. and Shaginyan, V. R.},
  journal = {JETP Lett.},
  volume = {51},
  pages = {553},
  year = {1990},
  doi = {},
  url = {http://jetpletters.ru/ps/0/article_17312.shtml}
}

@article{Kopnin2011,
  title = {High-temperature surface superconductivity in topological flat-band systems},
  author = {Kopnin, N. B. and Heikkil\"a, T. T. and Volovik, G. E.},
  journal = {Phys. Rev. B},
  volume = {83},
  pages = {220503(R)},
  year = {2011},
  doi = {10.1103/PhysRevB.83.220503}
}

@article{Gantmakher2010,
  title = {Superconductor–insulator quantum phase transition},
  author = {Gantmakher, Vsevolod F. and Dolgopolov, Valery T.},
  journal = {Phys. Usp.},
  volume = {53},
  pages = {1},
  year = {2010},
  doi = {10.3367/UFNe.0180.201001a.0003}
}

@book{SITbook,
  editor = {Dobrosavljevi\'{c}, Vladimir and Trivedi, Nandini and {Valles Jr.}, James M.},
  title = {Conductor-Insulator Quantum Phase Transitions},
  publisher = {Oxford University Press, Oxford},
  year = {2012},
  doi = {10.1093/acprof:oso/9780199592593.001.0001}
}

@article{Sacepe2020,
  title = {Quantum breakdown of superconductivity in low-dimensional materials},
  author = {Sac\'ep\'e, Benjamin and Feigel'man, Mikhail and Klapwijk, Tenunis M.},
  journal = {Nat. Phys.},
  volume = {16},
  pages = {734},
  year = {2020},
  doi = {10.1038/s41567-020-0905-x}
}

@unpublished{Yu2025,
  title = {Quantum Geometry in Quantum Materials},
  author = {Yu, Jiabin and Bernevig, B. Andrei and Queiroz, Raquel and Rossi, Enrico and T\"{o}rm\"{a}, P\"{a}ivi and Yang, Bohm-Jung},
  note = {},
  eprint = {2501.00098},
  archivePrefix = "arXiv"
}

@inproceedings{Peotta:ENFI25,
  author = {Peotta, Sebastiano and Huhtinen, Kukka-Emilia and T\"orm\"a, P\"aivi},
  title = {Quantum geometry in superfluidity and superconductivity},
  crossref = {enfi25},
  pages = {373--404},
  doi = {10.3254/ENFI250023}
}

@proceedings{enfi25,
  editor = {Rudolph Grimm and Massimo Inguscio and Giacomo Lamporesi and Sandro Stringari},
  title = {Proceedings of the International School of Physics ``Enrico Fermi"},
  booktitle = {Proceedings of the International School of Physics ``Enrico Fermi"},
  volume = {211},
  publisher = {IOP Press},
  year = {2025},
  isbn = {978-1-64368-580-9}
}

@article{Rossi2021,
  title = {Quantum metric and correlated states in two-dimensional systems},
  author = {Enrico Rossi},
  journal = {Curr. Opin. Solid State and Mater. Sci.},
  volume = {25},
  pages = {100952},
  year = {2021},
  doi = {10.1016/j.cossms.2021.100952}
}

@article{Chen2024,
  title = {Ginzburg-Landau Theory of Flat-Band Superconductors with Quantum Metric},
  author = {Chen, Shuai A. and Law, K. T.},
  journal = {Phys. Rev. Lett.},
  volume = {132},
  pages = {026002},
  year = {2024},
  doi = {10.1103/PhysRevLett.132.026002}
}

@article{Anderson1959,
  title = {Theory of dirty superconductors},
  author = {Anderson, P. W.},
  journal = {J. Phys. Chem. Solids},
  volume = {11},
  pages = {26-30},
  year = {1959},
  doi = {10.1016/0022-3697(59)90036-8}
}

@article{Anderson1958,
  title = {Absence of Diffusion in Certain Random Lattices},
  author = {Anderson, P. W.},
  journal = {Phys. Rev.},
  volume = {109},
  pages = {1492--1505},
  year = {1958},
  doi = {10.1103/PhysRev.109.1492}
}

@article{Abrikosov1959,
  title = {Superconducting alloys at finite temperatures},
  author = {Abrikosov, A. A. and Gor'kov, L. P.},
  journal = {Sov. Phys. JEPT},
  volume = {9},
  pages = {220--221},
  year = {1959}
}

@article{Bouadim2011,
  title = {Single- and two-particle energy gaps across the disorder-driven superconductor–insulator transition},
  author = {Bouadim, Karim and Loh, Yen Lee and Randeria, Mohit and Trivedi, Nandini},
  journal = {Nat. Phys.},
  volume = {7},
  pages = {884},
  year = {2011},
  doi = {10.1038/nphys2037}
}

@article{Poduval2022,
  title = {Subgap two-particle spectral weight in disordered $s$-wave superconductors: Insights from mode coupling approach},
  author = {Poduval, Prathyush P. and Samanta, Abhisek and Gupta, Prashant and Trivedi, Nandini and Sensarma, Rajdeep},
  journal = {Phys. Rev. B},
  volume = {106},
  pages = {064512},
  year = {2022},
  doi = {10.1103/PhysRevB.106.064512}
}

@article{Tasaki1998,
  title={From {Nagaoka}'s ferromagnetism to flat-band ferromagnetism and beyond: An introduction to ferromagnetism in the {Hubbard} model},
  author={Tasaki, Hal},
  journal={Prog. Theor. Phys.},
  volume={99},
  number={4},
  pages={489--548},
  year={1998},
  publisher={Oxford University Press},
  doi={10.1143/PTP.99.489}
}

@article{Leykam2018,
  title={Artificial flat band systems: from lattice models to experiments},
  author={Leykam, Daniel and Andreanov, Alexei and Flach, Sergej},
  journal={Adv. Phys.: X},
  volume={3},
  number={1},
  pages={1473052},
  year={2018},
  publisher={Taylor \& Francis},
  doi={10.1080/23746149.2018.1473052}
}

@article{Peotta2015,
  title={Superfluidity in topologically nontrivial flat bands},
  author={Peotta, Sebastiano and T{\"o}rm{\"a}, P{\"a}ivi},
  journal={Nat. Commun.},
  volume={6},
  number={1},
  pages={8944},
  year={2015},
  publisher={Nature Publishing Group UK London},
  doi={10.1038/ncomms9944}
}

@article{Julku2016,
  title={Geometric origin of superfluidity in the {Lieb}-lattice flat band},
  author={Julku, Aleksi and Peotta, Sebastiano and Vanhala, Tuomas I. and Kim, Dong-Hee and T{\"o}rm{\"a}, P{\"a}ivi},
  journal={Phys. Rev. Lett.},
  volume={117},
  number={4},
  pages={045303},
  year={2016},
  publisher={APS},
  doi={10.1103/PhysRevLett.117.045303}
}

@article{Liang2017,
  title={Band geometry, Berry curvature, and superfluid weight},
  author={Liang, Long and Vanhala, Tuomas I and Peotta, Sebastiano and Siro, Topi and Harju, Ari and T{\"o}rm{\"a}, P{\"a}ivi},
  journal={Phys. Rev. B},
  volume={95},
  number={2},
  pages={024515},
  year={2017},
  publisher={APS},
  doi={10.1103/PhysRevB.95.024515}
}

@article{Torma2018,
  title={Quantum metric and effective mass of a two-body bound state in a flat band},
  author={T{\"o}rm{\"a}, P{\"a}ivi and Liang, Long and Peotta, Sebastiano},
  journal={Phys. Rev. B},
  volume={98},
  number={22},
  pages={220511},
  year={2018},
  publisher={APS},
  doi={10.1103/PhysRevB.98.220511}
}

@article{Iskin2019,
  title={Origin of flat-band superfluidity on the Mielke checkerboard lattice},
  author={Iskin, M},
  journal={Phys. Rev. A},
  volume={99},
  number={5},
  pages={053608},
  year={2019},
  publisher={APS},
  doi={10.1103/PhysRevA.99.053608}
}

@article{Huhtinen2021,
  title = {Possible insulator-pseudogap crossover in the attractive Hubbard model on the Lieb lattice},
  author = {Huhtinen, Kukka-Emilia and T\"orm\"a, P\"aivi},
  journal = {Phys. Rev. B},
  volume = {103},
  pages = {L220502},
  year = {2021},
  doi = {10.1103/PhysRevB.103.L220502}
}

@article{Huhtinen2022,
  title={Revisiting flat band superconductivity: Dependence on minimal quantum metric and band touchings},
  author={Huhtinen, Kukka-Emilia and Herzog-Arbeitman, Jonah and Chew, Aaron and Bernevig, Bogdan A and T{\"o}rm{\"a}, P{\"a}ivi},
  journal={Phys. Rev. B},
  volume={106},
  pages={014518},
  year={2022},
  doi={10.1103/PhysRevB.106.014518}
}

@article{Ghosal1998,
  title={Role of spatial amplitude fluctuations in highly disordered s-wave superconductors},
  author={Ghosal, Amit and Randeria, Mohit and Trivedi, Nandini},
  journal={Phys. Rev. Lett.},
  volume={81},
  number={18},
  pages={3940},
  year={1998},
  publisher={APS},
  doi={10.1103/PhysRevLett.81.3940}
}

@article{Ghosal2001,
  title={Inhomogeneous pairing in highly disordered s-wave superconductors},
  author={Ghosal, Amit and Randeria, Mohit and Trivedi, Nandini},
  journal={Phys. Rev. B},
  volume={65},
  number={1},
  pages={014501},
  year={2001},
  publisher={APS},
  doi={10.1103/PhysRevB.65.014501}
}

@article{Ma1985,
  title={Localized superconductors},
  author={Ma, Michael and Lee, Patrick A},
  journal={Phys. Rev. B},
  volume={32},
  number={9},
  pages={5658},
  year={1985},
  publisher={APS},
  doi={10.1103/PhysRevB.32.5658}
}

@article{Dubi2007,
  title={Nature of the superconductor--insulator transition in disordered superconductors},
  author={Dubi, Yonatan and Meir, Yigal and Avishai, Yshai},
  journal={Nature},
  volume={449},
  number={7164},
  pages={876--880},
  year={2007},
  publisher={Nature Publishing Group UK London},
  doi={10.1038/nature06180}
}

@article{Scalettar1999,
  title={Quantum {Monte} {Carlo} study of the disordered attractive {Hubbard} model},
  author={Scalettar, RT and Trivedi, N and Huscroft, C},
  journal={Phys. Rev. B},
  volume={59},
  number={6},
  pages={4364},
  year={1999},
  publisher={APS},
  doi={10.1103/PhysRevB.59.4364}
}

@article{Trivedi1996,
  title={Superconductor-insulator transition in a disordered electronic system},
  author={Trivedi, Nandini and Scalettar, Richard T and Randeria, Mohit},
  journal={Phys. Rev. B},
  volume={54},
  number={6},
  pages={R3756},
  year={1996},
  publisher={APS},
  doi={10.1103/PhysRevB.54.R3756}
}

@article{Noda2009,
  title = {Ferromagnetism of cold fermions loaded into a decorated square lattice},
  author = {Noda, Kazuto and Koga, Akihisa and Kawakami, Norio and Pruschke, Thomas},
  journal = {Phys. Rev. A},
  volume = {80},
  pages = {063622},
  year = {2009},
  doi = {10.1103/PhysRevA.80.063622}
}

@article{Sakaida2013,
  title={Effects of Disorder on Superfluidity in the Attractive {Hubbard} Model},
  author={Sakaida, Masaru and Noda, Kazuto and Kawakami, Norio},
  journal={J. Phys. Soc. Jpn.},
  volume={82},
  number={7},
  pages={074715},
  year={2013},
  publisher={The Physical Society of Japan},
  doi={10.7566/JPSJ.82.074715}
}

@article{Iglovikov2014,
  title = {Superconducting transitions in flat-band systems},
  author = {Iglovikov, V. I. and H\'ebert, F. and Gr\'emaud, B. and Batrouni, G. G. and Scalettar, R. T.},
  journal = {Phys. Rev. B},
  volume = {90},
  pages = {094506},
  year = {2014},
  doi = {10.1103/PhysRevB.90.094506},
}

@article{Lau2022,
  title={Universal suppression of superfluid weight by non-magnetic disorder in s-wave superconductors independent of quantum geometry and band dispersion},
  author={Lau, Alexander and Peotta, Sebastiano and Pikulin, Dmitry and Rossi, Enrico and Hyart, Timo},
  journal={SciPost Phys.},
  volume={13},
  number={4},
  pages={086},
  year={2022},
  doi={10.21468/SciPostPhys.13.4.086}
}

@article{Kiran2024,
  title={Effect of correlated disorder on superconductivity in a kagome lattice: A {Bogoliubov}--de {Gennes} analysis},
  author={Kiran, Ravi and Biswas, Sudipta and Chakraborty, Monodeep},
  journal={Phys. Rev. B},
  volume={110},
  number={18},
  pages={184506},
  year={2024},
  publisher={APS},
  doi={10.1103/PhysRevB.110.184506}
}

@article{Bilitewski2018,
  title={Disordered flat bands on the kagome lattice},
  author={Bilitewski, Thomas and Moessner, Roderich},
  journal={Phys. Rev. B},
  volume={98},
  number={23},
  pages={235109},
  year={2018},
  publisher={APS},
  doi={10.1103/PhysRevB.98.235109}
}

@article{Tamura2002,
  title={Flat-band ferromagnetism in quantum dot superlattices},
  author={Tamura, Hiroyuki and Shiraishi, Kenji and Kimura, Takashi and Takayanagi, Hideaki},
  journal={Phys. Rev. B},
  volume={65},
  pages={085324},
  year={2002},
  doi={10.1103/PhysRevB.65.085324}
}

@article{Scalapino1993,
  title={Insulator, metal, or superconductor: The criteria},
  author={Scalapino, Douglas J and White, Steven R and Zhang, Shoucheng},
  journal={Phys. Rev. B},
  volume={47},
  number={13},
  pages={7995},
  year={1993},
  publisher={APS},
  doi={10.1103/PhysRevB.47.7995}
}

@article{Oliveira-Lima2020,
  title={Dynamical resilience to disorder: The dilute {Hubbard} model on the {Lieb} lattice},
  author={Oliveira-Lima, L and Costa, NC and de Lima, J Pimentel and Scalettar, RT and Santos, RR dos},
  journal={Phys. Rev. B},
  volume={101},
  pages={165109},
  year={2020},
  doi={10.1103/PhysRevB.101.165109}
}

@article{Li2022,
  title = {Metal-insulator transition in the disordered Hubbard model of the {Lieb} lattice},
  author = {Li, Yueqi and Tian, Lingyu and Ma, Tianxing and Lin, Hai-Qing},
  journal = {Phys. Rev. B},
  volume = {106},
  pages = {205149},
  year = {2022},
  doi = {10.1103/PhysRevB.106.205149}
}

@article{Chalker2010,
  title={Anderson localization in tight-binding models with flat bands},
  author={Chalker, JT and Pickles, TS and Shukla, Pragya},
  journal={Phys. Rev. B},
  volume={82},
  number={10},
  pages={104209},
  year={2010},
  publisher={APS},
  doi={10.1103/PhysRevB.82.104209}
}

@article{Roy2020,
  title = {Interplay of disorder and interactions in a flat-band supporting diamond chain},
  author = {Roy, Nilanjan and Ramachandran, Ajith and Sharma, Auditya},
  journal = {Phys. Rev. Res.},
  volume = {2},
  pages = {043395},
  year = {2020},
  doi = {10.1103/PhysRevResearch.2.043395}
}

@article{Bera2015,
  title={Many-body localization characterized from a one-particle perspective},
  author={Bera, Soumya and Schomerus, Henning and Heidrich-Meisner, Fabian and Bardarson, Jens H},
  journal={Phys. Rev. Lett.},
  volume={115},
  number={4},
  pages={046603},
  year={2015},
  publisher={APS},
  doi={10.1103/PhysRevLett.115.046603}
}

@article{Hopjan2021,
  title={Scaling properties of a spatial one-particle density-matrix entropy in many-body localized systems},
  author={Hopjan, Miroslav and Heidrich-Meisner, Fabian and Alba, Vincenzo},
  journal={Phys. Rev. B},
  volume={104},
  number={3},
  pages={035129},
  year={2021},
  publisher={APS},
  doi={10.1103/PhysRevB.104.035129}
}

@article{Bera2017,
  title={One-particle density matrix characterization of many-body localization},
  author={Bera, Soumya and Martynec, Thomas and Schomerus, Henning and Heidrich-Meisner, Fabian and Bardarson, Jens H},
  journal={Ann. Phys.},
  volume={529},
  number={7},
  pages={1600356},
  year={2017},
  publisher={Wiley Online Library},
  doi={10.1002/andp.201600356}
}

@article{Orito2021,
  title={Multifractality and Fock-space localization in many-body localized states: One-particle density matrix perspective},
  author={Orito, Takahiro and Imura, Ken-Ichiro},
  journal={Phys. Rev. B},
  volume={103},
  number={21},
  pages={214206},
  year={2021},
  publisher={APS},
  doi={10.1103/PhysRevB.103.214206}
}

@article{Liang2023,
  title={Disorder in interacting quasi-one-dimensional systems: Flat and dispersive bands},
  author={Liang, Mi-Ji and Yang, Yong-Feng and Cheng, Chen and Mondaini, Rubem},
  journal={Phys. Rev. B},
  volume={108},
  pages={035131},
  year={2023},
  doi={10.1103/PhysRevB.108.035131}
}

@unpublished{Lebrat2025,
  title = {Ferrimagnetism of ultracold fermions in a multi-band {Hubbard} system},
  author = {Lebrat, Martin and Kale, Anant and Kendrick, Lev Haldar and Xu, Muqing and Gang, Youqi and Nikolaenko, Alexander and Bonetti, Pietro M. and Sachdev, Subir and Greiner, Markus},
  note = {},
  eprint = {2404.17555},
  archivePrefix = "arXiv"
}

@article{Tam2024,
  title = {Geometry-independent superfluid weight in multiorbital lattices from the generalized random phase approximation},
  author = {Tam, Minh and Peotta, Sebastiano},
  journal = {Phys. Rev. Res.},
  volume = {6},
  pages = {013256},
  year = {2024},
  doi = {10.1103/PhysRevResearch.6.013256}
}

@article{Goda2006,
  title = {Inverse {Anderson} Transition Caused by Flatbands},
  author = {Goda, Masaki and Nishino, Shinya and Matsuda, Hiroki},
  journal = {Phys. Rev. Lett.},
  volume = {96},
  pages = {126401},
  year = {2006},
  doi = {10.1103/PhysRevLett.96.126401}
}

@article{Nishino2007,
  author = {Nishino, Shinya and Matsuda, Hiroki and Goda, Masaki},
  title = {Flat-Band Localization in Weakly Disordered System},
  journal = {J. Phys. Soc. Jpn.},
  volume = {76},
  pages = {024709},
  year = {2007},
  doi = {10.1143/JPSJ.76.024709}
}

@article{Leykam2013,
  title = {Flat band states: Disorder and nonlinearity},
  author = {Leykam, Daniel and Flach, Sergej and Bahat-Treidel, Omri and Desyatnikov, Anton S.},
  journal = {Phys. Rev. B},
  volume = {88},
  pages = {224203},
  year = {2013},
  doi = {10.1103/PhysRevB.88.224203}
}

@article{Leykam2017,
  title = {Localization of weakly disordered flat band states},
  author = {Leykam, Daniel and Bodyfelt, Joshua D. and Desyatnikov, Anton S. and Flach, Sergej},
  journal = {Eur. Phys. J. B},
  volume = {90},
  pages = {1},
  year = {2017},
  doi = {10.1140/epjb/e2016-70551-2}
}

@article{Shukla2018a,
  title = {Disorder perturbed flat bands: Level density and inverse participation ratio},
  author = {Shukla, Pragya},
  journal = {Phys. Rev. B},
  volume = {98},
  pages = {054206},
  year = {2018},
  doi = {10.1103/PhysRevB.98.054206}
}

@article{Shukla2018b,
  title = {Disorder perturbed flat bands. {II}. Search for criticality},
  author = {Shukla, Pragya},
  journal = {Phys. Rev. B},
  volume = {98},
  pages = {184202},
  year = {2018},
  doi = {10.1103/PhysRevB.98.184202}
}

@article{Schirmer2025,
  title = {Superfluid weight of strongly inhomogeneous superconductors},
  author = {Schirmer, Jonathan and Rossi, Enrico},
  journal = {Phys. Rev. B},
  volume = {112},
  pages = {L180501},
  year = {2025},
  doi = {10.1103/9gc8-q16z}
}

@article{Penrose1956,
  title = {Bose-Einstein Condensation and Liquid Helium},
  author = {Penrose, Oliver and Onsager, Lars},
  journal = {Phys. Rev.},
  volume = {104},
  pages = {576--584},
  year = {1956},
  doi = {10.1103/PhysRev.104.576}
}

@article{Rigol2004,
  title = {Emergence of Quasicondensates of Hard-Core Bosons at Finite Momentum},
  author = {Rigol, Marcos and Muramatsu, Alejandro},
  journal = {Phys. Rev. Lett.},
  volume = {93},
  pages = {230404},
  year = {2004},
  doi = {10.1103/PhysRevLett.93.230404}
}

@article{Rigol2005,
  title = {Finite-temperature properties of hard-core bosons confined on one-dimensional optical lattices},
  author = {Rigol, Marcos},
  journal = {Phys. Rev. A},
  volume = {72},
  pages = {063607},
  year = {2005},
  doi = {10.1103/PhysRevA.72.063607}
}

@article{Nessi2011,
  title = {Finite-temperature properties of one-dimensional hard-core bosons in a quasiperiodic optical lattice},
  author = {Nessi, Nicolas and Iucci, An\'{\i}bal},
  journal = {Phys. Rev. A},
  volume = {84},
  pages = {063614},
  year = {2011},
  doi = {10.1103/PhysRevA.84.063614}
}

@article{Gramsch2012,
  title = {Quenches in a quasidisordered integrable lattice system: Dynamics and statistical description of observables after relaxation},
  author = {Gramsch, Christian and Rigol, Marcos},
  journal = {Phys. Rev. A},
  volume = {86},
  pages = {053615},
  year = {2012},
  doi = {10.1103/PhysRevA.86.053615}
}

@article{Chan2022,
  title = {Designer flat bands: Topology and enhancement of superconductivity},
  author = {Chan, Si Min and Gr\'emaud, B. and Batrouni, G. G.},
  journal = {Phys. Rev. B},
  volume = {106},
  pages = {104514},
  year = {2022},
  doi = {10.1103/PhysRevB.106.104514}
}

@article{Wu2021,
  title = {Superfluid density and collective modes of fermion superfluid in dice lattice},
  author = {Wu, Yu-Rong and Zhang, Xiao-Fei and Liu, Chao-Fei and Liu, Wu-Ming and Zhang, Yi-Cai},
  journal = {Sci. Rep.},
  volume = {11},
  pages = {13572},
  year = {2021},
  doi = {https://doi.org/10.1038/s41598-021-93007-z}
}

@article{Bergman2008,
  title = {Band touching from real-space topology in frustrated hopping models},
  author = {Bergman, Doron L. and Wu, Congjun and Balents, Leon},
  journal = {Phys. Rev. B},
  volume = {78},
  pages = {125104},
  year = {2008},
  doi = {10.1103/PhysRevB.78.125104}
}

@article{Rhim2019,
  title = {Classification of flat bands according to the band-crossing singularity of Bloch wave functions},
  author = {Rhim, Jun-Won and Yang, Bohm-Jung},
  journal = {Phys. Rev. B},
  volume = {99},
  pages = {045107},
  year = {2019},
  doi = {10.1103/PhysRevB.99.045107}
}

@article{Rhim2021,
  author = {Jun-Won Rhim and Bohm-Jung Yang},
  title = {Singular flat bands},
  journal = {Adv.Phy. X},
  volume = {6},
  pages = {1901606},
  year = {2021},
  doi = {10.1080/23746149.2021.1901606}
}

@unpublished{Kolar2025,
  title={Superfluid weight in disordered flat-band superconductors as a competition between localization functionals}, 
  author={Kry\v{s}tof Kol\'{a}\v{r} and Tero T. Heikkil\"{a} and P\"{a}ivi T\"{o}rm\"{a}},
  note = {},
  eprint = {2510.05224},
  archivePrefix = "arXiv"
}

@article{Chen2025,
  title = {Generalized {Peierls} substitution for {Wannier} obstructions: Response to disorder and interactions},
  author = {Chen, Shuai A. and Moessner, Roderich and Ng, Tai Kai},
  journal = {Phys. Rev. Lett.},
  volume = {135},
  pages = {116502},
  year = {2025},
  doi = {10.1103/xkxw-1134}
}

@article{Chan2025,
  title = {Disorder and robustness of superconductivity on the quasi-one-dimensional flat-band {Creutz} lattice},
  author = {Chan, Si Min and Gr\'emaud, B. and Batrouni, G. George},
  journal = {Phys. Rev. B},
  volume = {112},
  pages = {094501},
  year = {2025},
  doi = {10.1103/nn9h-r25c}
}

@article{Bouzerar2025,
  title = {Robustness of flat band superconductivity against disorder in a two-dimensional {Lieb} lattice model},
  author = {Bouzerar, G. and Thumin, M.},
  journal = {Phys. Rev. B},
  volume = {111},
  pages = {L020506},
  year = {2025},
  doi = {10.1103/PhysRevB.111.L020506}
}

\end{document}